\documentclass[preprint,aps]{revtex4}
\begin{document}
\draft
\title{Q -- phonon description of low lying 1$^-$ two -- phonon states
in spherical nuclei}
\author{R.V.Jolos$^{a}$, N.Yu.Shirikova$^{a}$, V.V.Voronov$^{a}$}
\affiliation{
$^{a}$Joint Institute for Nuclear Research, 141980 Dubna, Russia\\}
\date{\today}
\begin{abstract}
The properties of $1^-_1$ two--phonon states and
the characteristics of E1 transition probabilities between
low--lying collective states in spherical nuclei are  analysed within
the $Q$--phonon approach to the description of collective states.
Several relations between observables are obtained. Microscopic
calculations of the E1 $0^+_1\rightarrow 1^-_1$ transition matrix elements are
performed on the basis of the RPA. A satisfactory description of the
experimental data is obtained.
\end{abstract}
\maketitle

\section{Introduction}

In systematic investigations of the E1 transitions
\cite{Guhr,Zilges,Kneissl,Fransen,Robinson,Wilhelm1,Wilhelm2}
the low--lying 1$^{-}_1$ have been observed in
spherical nuclei
states, which are characterized by strong B(E1;$0^+_{1}\rightarrow 1^{-}_1 $), of
the order of several units$\times 10^{-3}e^2 fm^2$.
It was demonstrated that these
low--lying
1$^{-}_1$ states arise as $|2^+_1\otimes 3^-_1 ;1^- M\rangle$ due to coupling of the collective quadrupole 2$^+_1$ and
the collective octupole
3$^-_1$ states. Similar 1$^-_1$
states
have been observed in the Cd, Sn, Ba, Ce, Nd, and Sm isotopes.
Their two--phonon
character has been proved by the observed strong E2 and E3 transitions to
the
corresponding  one--phonon states  \cite{Robinson,Wilhelm1,Wilhelm2}
and by the fact that the
energies of these
1$^-_1$  states are very close to the summed energies $E(2^+_1) + E(3^-_1)$.

Strong correlations between the values of B(E1;$0^+_{1}\rightarrow 1^-_1 $)
and the
product of average squares of the quadrupole $\langle \beta^2_2\rangle$
and octupole  $\langle \beta^2_3\rangle$ deformation parameters in nonmagic nuclei
\cite{Babilon} clearly show that away from the closed shells
the large B(E1) values
have a collective nature connected with the motion of the nuclear shape. The
ratio $B(E1)$/$(\langle \beta^2_3\rangle\langle \beta^2_3\rangle )$ is amazingly
constant, although the B(E1) strength varies by an order of magnitude in the
considered nuclei.

Accepting a two--phonon picture of the 1$^-_1$ states let us consider the
other important
experimental facts characterizing strong E1 transitions between low--lying states.\\
-- Analyzing the experimental data on the ratio
B(E1;$1^-_1\rightarrow 0^+_{1})$/B(E1;$3^-_1\rightarrow 2^+_1)$
it was shown in
\cite{Pietralla1} that this ratio is a constant equal to 1 within a factor of 2
although these E1 transition strengths can differ by about two orders of
magnitude for different nuclei. Semimagic even--$A$ nuclei and even--even nuclei
with two or four nucleons outside a closed shell were considered in this
analysis. These nuclei are rather vibrational and can be understood, at least qualitatively,
in a harmonic phonon picture. In the last model
B(E1;$1^-_1\rightarrow 0^+_{1})$/B(E1;$3^-_1\rightarrow 2^+_1)=\frac{7}{3}$.
This correlation of the E1 strengths can be considered as an additional support
for the quadrupole -- octupole coupled character of the 1$^-_1$ states.\\
-- The B(E1;$1^-_1\rightarrow 2^+_1)$ is small in Sn isotopes; however, it increase
in Cd, Te isotopes and approach the Alaga value for the ratio
B(E1;$1^-_1\rightarrow 2^+_1)$/B(E1;$1^-_1\rightarrow 0^+_{1})$
in deformed nuclei \cite{Andrejtscheff}.\\
-- A minimum has been found in the $A$--dependence of B(E1;$1^-_1\rightarrow 0^+_{1})$
in the Nd, Sm and Ba isotopes when the number of
neutrons $N$ is equal to 78 or 86 \cite{Metzger1,Metzger2,Eckert}.
This characteristic feature of the behavior of B(E1;$1^-_1\rightarrow 0^+_{1})$ as a function of $N$
has been discovered earlier in the RPA--based calculations in \cite{Voronov}.\\
-- Strong E1 transitions between the $3^-_1$ octupole--phonon states and the
2$^+_{1,ms}$ mixed--symmetry quadrupole one--phonon states in the spherical
nuclei  $^{92}_{40}$Zr$_{52}$, $^{94}_{42}$Mo$_{52}$, $^{96}_{44}$Ru$_{52}$  and
$^{142}_{58}$Ce$_{84}$ are observed \cite{Pietralla5}.
In these four nuclei the E1 transitions to
the 2$^+_{1,ms}$ state are stronger than to the 2$^+_{1}$ state. Probably, this is
a consequence of the isovector nature of the E1 transition operator,
which enhances E1 transitions between the mixed symmetry states and the isoscalar ones.
In the last case, this is the isoscalar octupole vibrational state.\\
Since B(E1;$3^-_1\rightarrow 2^+_{ms})$ are strong the question arises if there should appear
strong E1 transitions from the ground state to the
$(3^-_1\otimes 2^+_{ms})_{1^-}$ state, which is expected at the excitation energy
4187 keV in $^{92}$Zr, 4601 keV in $^{94}$Mo, 4934 keV in $^{96}$Ru and
3658 keV in $^{142}$Ce. However, anharmonic effects, which can
shift the energies of
these two -- phonon states, should not be excluded.

Summarizing the short review of the data we can see that the microscopical model
describing strong E1 transitions at low energies basing on the two--phonon
quadrupole--octupole model of the low lying 1$^-_1$ states should explain:\\
-- the fact that B(E1;$1^-_1\rightarrow 0^+_{1})$/B(E1;$3^-_1\rightarrow 2^+_1)\approx 1$
at least in near magic nuclei;\\
--an increase of B(E1;$1^-_1\rightarrow 2^+_1)$ when going away from the
closed shell;\\
-- the existence of the minimum in the $A$ -- dependence of B(E1;$0^+_{1}\rightarrow 1^-_1)$
when moving off the semimagic nuclei;\\
-- the fact that B(E1;$3^-_1\rightarrow 2^+_{ms})\gg$ B(E1;$3^-_1\rightarrow 2^+_1)$.\\
-- proportionality of B(E1;$0^+_{1}\rightarrow 1^-_1)$ to
$(\langle \beta^2_2\rangle\langle \beta^2_3\rangle )$ in collective nuclei;\\

\section{Model}

In \cite{Pietralla2,Pietralla3,Palchikov} the $Q$ -- phonon approach
to the description of the positive parity collective states was developed within the Interacting Boson Model.
It was shown that the wave vectors
of the yrast, the second $2^+$ and the second $0^+$ states could be described to a high accuracy over the whole parameter space
of the consistent $Q$ Hamiltonian, i.e., far outside of the region  where a picture of harmonic vibrations
was correct, by simple universal expressions contaning only one or two multiple $Q$--phonon configurations.
A simple structure of the wave vectors helps derive different relations between transition matrix elements.

In contrast to \cite{Pietralla2,Pietralla3,Palchikov}, where the $Q$--phonon approach was formulated for the
bosonic configurational space of the Interacting Boson Model, in the present paper this approach is formulated
for the fermionic configurational space. In addition, we consider both positive and negative parity
states.

In the $Q$--phonon approach the $2^+_1$ state is presented by the following expression
\begin{equation}
\label{1-eq}
|2^+_1,\mu\rangle ={\cal N}_{2^+_1}\hat{Q}_{2\mu}|0^+_1\rangle ,
\end{equation}
where $|0^+_1\rangle$ is the ground state vector,
${\cal N}_{2^+_1}=\left(\frac{1}{\sqrt{5}}\langle 0^+_1|(\hat{Q}_2\hat{Q}_2)_0|0^+_1\rangle \right)^{-1/2}$\\ = $\sqrt{5}$/ $\mid\langle0^+_1\parallel Q_2\parallel 2^+_1\rangle\mid$
and $\hat{Q}_{2\mu}$ is in our case the standard shell model quadrupole moment operator
\begin{equation}
\label{2-eq}
\hat{Q}_{2\mu}=\sum_{jj'mm'} \langle jm|r^2 Y_{2\mu} |j'm'\rangle a^+_{jm}a_{j'm'},
\end{equation}
expressed in terms of the nucleon creation $a^+_{jm}$ and annihilation $a_{j'm'},$ operators.

Let us
include into consideration the octupole mode
\begin{equation}
\label{3-eq}
|3^-_1,\nu\rangle ={\cal N}_{3^-_1}\hat{Q}_{3\nu}|0^+_1\rangle ,
\end{equation}
where ${\cal N}_{3^-_1}=\left(\frac{1}{\sqrt{7}}\langle 0^+_1|(\hat{Q}_3\hat{Q}_3)_0|0^+_1\rangle \right)^{-1/2}$ = $\sqrt{7}$/ $\mid\langle0^+_1\parallel Q_3\parallel 3^-_1\rangle\mid$
and $\hat{Q}_{3\nu}$ is the fermionic octupole moment operator
\begin{equation}
\label{4-eq}
\hat{Q}_{3\nu}=\sum_{jj'mm'} \langle jm|r^3 Y_{3\mu} |j'm'\rangle a^+_{jm}a_{j'm'}.
\end{equation}
As we know from the RPA type calculations \cite{Soloviev} both expressions (\ref{1-eq}) and (\ref{3-eq})
 are very good approximations for the lowest collective $2^+$ and $3^-$ states. In fact, this is
 a consequence of the well--known from the experiment result that  E2 and  E3 transitions from the ground state
 to the $2^+_1$ and $3^-_1$ states, correspondingly, are much stronger than transitions to
 higher lying $2^+$ and $3^-$ states.

Continuing along this line and remembering that the 1$^-_1$ state is mainly a quadrupole -- octupole
two -- phonon state we suggest for the 1$^-_1$ state vector that
\begin{equation}
\label{5-eq}
|1^-_1, M\rangle ={\cal N}_{1^-_1}\left(\hat{Q}_{2}\hat{Q}_{3}\right)_{1M}|0^+_1\rangle ,
\end{equation}
where for ${\cal N}_{1^-_1}$ we obtain the following relation:
\begin{eqnarray}
\label{6-eq}
\left({\cal N}_{1^-_1}\right)^{-2}&=&\left({\cal N}_{2^+_1}\right)^{-2}\left({\cal N}_{3^-_1}\right)^{-2}\nonumber\\
&+& \frac{1}{\sqrt{5}}\sum_{n\ne 1}\langle 0^+_1 |(Q_2 Q_2)_0 |0^+_n\rangle\langle 0^+_n |(Q_3 Q_3)_0 |0^+_1\rangle\nonumber\\
&+&\frac{2\sqrt{30}}{35}\langle 0^+_1 |\left( (Q_2 Q_2)_2 (Q_3 Q_3)_2\right)_0 |0^+_1\rangle\nonumber\\
&+&\frac{\sqrt{11}}{7\sqrt{5}}\langle 0^+_1 |\left( (Q_2 Q_2)_4 (Q_3 Q_3)_4\right)_0 |0^+_1\rangle
\end{eqnarray}
The last three terms in the right--hand side of (\ref{6-eq}) can give a noticeable contribution
only in the case of strong mixing of the two--phonon quadrupole--quadrupole and two--phonon
octupole--octupole states. However, as it is known from the RPA type calculations \cite{Grinberg}, this
mixing is insignificant: less than 1$\%$. Neglecting these terms we obtain an approximate relation
\begin{equation}
\label{7-eq}
{\cal N}_{1^-_1}\approx {\cal N}_{2^+_1}{\cal N}_{3^-_1}.
\end{equation}

The wave vector (\ref{5-eq}) can be written in terms of the RPA collective phonons and the two--quasiparticle
components which correspond to the noncollective RPA solutions. Written in this way the wave vector (\ref{5-eq})
has the two--phonon component with one quadrupole and one octupole collective RPA phonons as the main component, the
three--phonon component with two collective quadrupole and one collective octupole RPA phonons, and
the two--quasiparticle $1^-$ components. It means that the results of our $Q$ -- phonon approach should
be compaired not with the pure RPA calculations but with the RPA based calculations which include also
anharmonic effects. A contribution of the last two components to the norm of the wave vector (\ref{5-eq}) is very
small for nuclei considered in the present paper.

Of course, when we approach a deformed region the following $Q$ -- phonon component
\begin{equation}
\label{8-eq}
\left (\left (Q_2 Q_2\right)_4 Q_3\right )_{1M}|0^+_1\rangle
\end{equation}
becomes more and more important since in this region there are not one but two $1^-$ states with $K$=0 and $K$=1.
To construct the $1^-$ state with $K$ as a good quantum number, it is necessary to take a mixture of the states
(\ref{5-eq}) and (\ref{8-eq}). This mixing, which can be estimated as 70$\%$ of one component plus 30$\%$
of the other one can be considered as an upper limit for the error connected with the approximation (\ref{5-eq}).
Of course, near spherical nuclei this mixing is much smaller. In transitional nuclei with soft $\gamma$ -- mode
the $Q$ -- phonon state
\begin{equation}
\label{9-eq}
\left (\left (Q_2 Q_2\right )_2 Q_3\right )_{1M}|0^+_1\rangle
\end{equation}
can also become important.

The strength of the $Q$ -- phonon scheme lies in a possibility to derive relations between electromagnetic transition matrix elements
also outside the analytically solvable harmonic vibrator and rotor limits. This is possible because of the
simple form of the wave vectors in this approach. This possibility is realized in this section below and in
sections III and IV. However, the wave vectors (\ref{1-eq}), (\ref{3-eq}) and (\ref{5-eq}) only look simple.
The multipole moment operators in these expressions act on the exact ground state. Thus, for calculations we
needed an expression for the ground state vector. In section V we take as an approximation
to the ground state vector the vacuum state of the RPA phonons.

Let us apply the model formulated above to calculations of  the matrix element
of the E1 $0^+_1\rightarrow 1^-_1$ transition. We
obtain
\begin{equation}
\label{10-eq}
\langle 1^-_1, M|{\cal M}_{1M}(E1)|0^+_{1}\rangle ={\cal N}_{1^-_1}\langle 0|(Q_2Q_3)_{1M}{\cal M}_{1M}(E1)|0^+_1\rangle ,
\end{equation}
where ${\cal M}_{1M}(E1)$ is the operator of the electric dipole transition.
Since multipole operators $Q_2$, $Q_3$ and ${\cal M}(E1)$ commute with each other we can rewrite the last expression as
\begin{equation}
\label{11-eq}
\langle 1^-_1, M|{\cal M}_{1M}(E1)|0^+_{1}\rangle ={\cal N}_{1^-_1}\sum_{\mu ,\nu} C^{1M}_{2\mu  3\nu}\langle 0^+_1|(-1)^{\nu} Q_{3-\nu}{\cal M}_{1M}(E1)(-1)^{\mu}Q_{2-\mu}|0^+_1\rangle
\end{equation}
Using expressions (\ref{1-eq}), (\ref{3-eq}) and relation (\ref{7-eq}) we can
rewrite (\ref{11-eq}) in the following way:
\begin{equation}
\label{12-eq}
\langle 1^-_1, M|{\cal M}_{1M}(E1,M)|0^+_{1}\rangle =\sum_{\mu ,\nu} C^{1M}_{2\mu  3\nu}(-1)^{\mu}\langle  3^-_1,\nu |{\cal M}_{1M}(E1)|2^+_1,-\mu\rangle
\end{equation}
>From the last expression using the Wigner -- Eckart theorem we obtain
\begin{equation}
\label{13-eq}
\langle 1^-_1\parallel {\cal M}(E1)\parallel 0^+_1\rangle = \langle 3^-_1\parallel {\cal M}(E1)\parallel 2^+_1\rangle ,
\end{equation}
and from (\ref{13-eq}) we get
\begin{equation}
\label{14-eq}
B(E1;1^-_1\rightarrow 0^+_{1})/B(E1;3^-_1\rightarrow 2^+_1)=\frac{7}{3}.
\end{equation}
The last value coincides with that of the harmonic phonon picture.
However, we did not use the assumption of harmonicity of
quadrupole and octupole vibrations.

Thus, the $Q$--phonon model explains
qualitatively one of the experimental results mentioned in Introduction.
However, the value of the ratio 7/3 deviates from the experimental
observation 1$\div$2. A possible reason for this deviation can be the presence
of some admixtures in the $|1^-_1\rangle$,
$|2^+_1\rangle$  and
$|3^-_1\rangle$ states which are not taken into account by the $Q$--phonon approach. The corrections to
relation (\ref{7-eq}) discussed above will also influence the number in the right--hand side of (\ref{14-eq}).
Mention, however, a very small value of $B(E1;3^-_1\rightarrow 2^+_1)$ obtain in \cite{Vanhoy}.
Taken together with the known value of $B(E1;1^-_1\rightarrow 0^+_{1})$ it produces the value of the ratio
(\ref{14-eq}) which is much larger than 7/3.

\section{E1 transition $1^-_1\rightarrow 2^+_1$}

Let us consider the matrix element $\langle 1^-_1 M'\mid {\cal M}_{1\mu}(E1)\mid 2^+_1 M\rangle$.
This matrix element can be written in the following way:
\begin{eqnarray}
\label{15-eq}
\langle 1^-_1 M'\mid {\cal M}_{1\mu}(E1)\mid 2^+_1 M\rangle ={\cal N}_{2^+_1}\langle 1^-_1 M'\mid {\cal M}_{1\mu}(E1)Q_{2M}\mid 0^+_1\rangle\nonumber\\
={\cal N}_{2^+_1}\langle 1^-_1 M'\mid Q_{2M} {\cal M}_{1\mu}(E1)\mid 0^+_1\rangle
\end{eqnarray}
Having in mind that the $0^+_1\rightarrow 1^-_1$ transition is the strongest one among the low--lying states
we can rewrite approximately (\ref{15-eq}) as
\begin{eqnarray}
\label{16-eq}
\langle 1^-_1 M'\mid {\cal M}_{1\mu}(E1)\mid 2^+_1 M\rangle &\approx& {\cal N}_{2^+_1}\langle 1^-_1 M'\mid Q_{2M}\mid1^-_1 \mu\rangle\langle 1^-_1 \mu\mid {\cal M}_{1\mu}(E1)\mid 0^+_1\rangle\nonumber\\
&=&{\cal N}_{2^+_1}\langle 1^-_1 M'\mid Q_{2M}\mid1^-_1 \mu\rangle\frac{1}{ \sqrt{3}}\langle 1^-_1\parallel {\cal M}(E1)\parallel 0^+_1\rangle\nonumber\\
&=&{\cal N}_{2^+_1}\frac{1}{3} C^{1M'}_{1\mu  2M}\langle 1^-_1\parallel Q_2\parallel 1^-_1\rangle\langle 1^-_1\parallel {\cal M}(E1)\parallel 0^+_1\rangle
\end{eqnarray}
and as a consequence of (\ref{16-eq}) we obtain
\begin{eqnarray}
\label{17-eq}
\frac{\langle 1^-_1\parallel {\cal M}(E1)\parallel  2^+_1\rangle}{\langle 1^-_1\parallel {\cal M}(E1)\parallel  0^+_1\rangle}=\sqrt{ \frac{5}{3}}\frac{\langle 1^-_1\parallel Q_2\parallel 1^-_1\rangle}{\langle 0^+_1\parallel Q_2\parallel 2^+_1\rangle}
\end{eqnarray}
The quadrupole moment of the $1^-_1$ state can be expressed approximately in terms of the quadrupole moments
of the $2^+_1$ and $3^-_1$ states.
Finally, we have
\begin{eqnarray}
\label{18-eq}
\frac{\langle 1^-_1\parallel {\cal M}(E1)\parallel  2^+_1\rangle}{\langle 1^-_1\parallel {\cal M}(E1)\parallel  0^+_1\rangle}\approx\frac{1}{\sqrt{35}}\frac{\langle 2^+_1\parallel Q_2\parallel 2^+_1\rangle}{\langle 0^+_1\parallel Q_2\parallel 2^+_1\rangle} + \frac{\sqrt{6}}{\sqrt{35}}\frac{\langle 3^-_1\parallel Q_2\parallel 3^-_1\rangle}{\langle 0^+_1\parallel Q_2\parallel 2^+_1\rangle}.
\end{eqnarray}
>From (\ref{18-eq}) we can see that going from the magic or semimagic nuclei to  nuclei with open shell the ratio
B(E1;$1^-_1\rightarrow 2^+_1$)/B(E1;$1^-_1\rightarrow 0^+_1$) increases with
the quadrupole moments of the first $2^+$ and $3^-$ states. In order to obtain
the Alaga rule for the ratio of the matrix elements
$\langle 1^-_1\parallel {\cal M}(E1)\parallel 2^+_1\rangle$/$\langle 1^-_1\parallel {\cal M}(E1)\parallel 0^+_1\rangle$,
we must take into account an admixture of $(|4^+_1\rangle\otimes |3^-_1\rangle )_1$
to the $|1^-_1\rangle$ state.

\section{IBM analysis of the neutron number dependence of the
B(E1; $0^+_1\rightarrow 1^-_1$)}

For completeness of the consideration the results of the very schematic IBM--based
analysis of the E1 transitions between the low--lying collective states are given in this section below.
A detailed analysis is presented in
\cite{Pietralla5,Smirnova}.
Let us analyse qualitatively the quantity B(E1;$0^+_1\rightarrow 1^-_1$) in order
to understand a possible reason for the appearance of the minimum in the neutron
number dependence of this quantity when  going away from the closed
neutron shell $N$=82.
As it follows from relation (\ref{13-eq}), instead of
$\langle 1^-_1\parallel Q(E1)\parallel 0^+_1\rangle$ we can consider the matrix
element $\langle 3^-_1\parallel Q(E1)\parallel 2^+_1\rangle$.
In the IBM $|2^+_1\rangle$ and $|3^-_1\rangle$ are the states
with the maximum value of the $F$--spin: $F=F_{max}$. The dipole transition operator
${\cal M}(E1)$ is mainly the $F$--spin vector. To simplify the discussion,  we assume below that the E1 transition
operator is exactly the $F$--spin vector. Then
\begin{equation}
\label{19-eq}
\langle 3^-_1\parallel Q(E1)\parallel  2^+_1\rangle\sim (N_{\pi} - N_{\nu}),
\end{equation}
where $2N_{\pi}$ and $2N_{\nu}$ are the numbers of the
valence protons and neutrons, respectively.
If we consider a nucleus with the closed neutron, shell, then as it is seen
from (\ref{19-eq}), only protons will contribute to the matrix element
$\langle 3^-_1\parallel Q(E1)\parallel 2^+_1\rangle$. If we start to increase the number
of valence neutrons, then a neutron contribution to
$\langle 3^-_1\parallel Q(E1)\parallel 2^+_1\rangle$ will increase compensating,
partly, a proton contribution. This matrix element takes the minimum value
when $N_{\pi} = N_{\nu}$. Of course, this phenomenological
analysis cannot give us the value of $N_{\pi}$ at which $\langle 3^-_1\parallel Q(E1)\parallel 2^+_1\rangle$
has a minimum because the physical E1 transition operator does not reduce to the $F$ -- spin vector.
In the IBM--based consideration it is assumed, in fact,   that
a microscopic structure of the proton and the neutron bosons does not change
from nucleus to nucleus. However, it is not evident that this assumption will be supported by the
microscopic calculations.
The results of the RPA based microscopic analysis are presented below.

In a similar way, transitions between the $3^-_1$ and $2^+_{ms}$
states \cite{Pietralla5} can be considered. In contrast to the preceding case the
$| 2^+_{ms}\rangle$ state is a state with $F=F_{max}-1$. Therefore,
\begin{eqnarray}
\label{20-eq}
\langle 3^-_1\parallel Q(E1)\parallel  2^+_{ms}\rangle&\sim& C^{F_{max}M_F}_{F_{max}-1  M_F\quad 10}\nonumber\\
&=&\sqrt{ \frac{(F_{max}+M_F)(F_{max}-M_F)}{(2F_{max}-1)F_{max}}},
\end{eqnarray}
where $F_{max}=\frac{1}{2}(N_{\pi}+N_{\nu})$ and $M_F=\frac{1}{2}(N_{\pi}-N_{\nu})$.
Therefore,
\begin{eqnarray}
\label{21-eq}
\langle 3^-_1\parallel Q(E1)\parallel  2^+_{ms}\rangle&\sim&\sqrt{ \frac{N_{\pi}N_{\nu}}{(2N_{\pi}+2N_{\nu}-1)(N_{\pi}+N_{\nu})}}.
\end{eqnarray}
It is seen from (\ref{20-eq}) and (\ref{21-eq}) that in contrast to
(\ref{19-eq}) there is no cancellation of the proton and neutron contributions
to the matrix element.
Thus, this analysis indicates a possible reason for a large value of the
$\langle 3^-_1\parallel Q(E1)\parallel  2^+_{ms}\rangle$ matrix element
\cite{Pietralla5}.

Let us assume, by analogy with  section II, that the $|2^+_{ms}\rangle$
state is created by
an action of the operator
\begin{eqnarray}
\label{mixed-eq}
\left(\frac{Q_2^{\nu}}{\langle 0^+_1 |(Q_2^{\nu}Q_2^{\nu})_0 |0^+_1\rangle^{1/2}}-\frac{Q_2^{\pi}}{\langle 0^+_1 |(Q_2^{\pi}Q_2^{\pi})_0 |0^+_1\rangle^{1/2}}\right)
\end{eqnarray}
on the ground
state. Here $Q_2^{\nu(\pi)}$ is the neutron (proton) quadrupole operator. This state is constructed
so as to be orthogonal to the $2^+_1$ state.
Then using as above a commutativity of the operators (\ref{mixed-eq})
and ${\cal M}(E1)$ we obtain from the inequality\\
$\langle 3^-_1\parallel{\cal M}(E1)\parallel 2^+_{ms}\rangle\gg \langle 3^-_1\parallel{\cal M}(E1)\parallel 2^+_1\rangle$
that
$\langle (3^-_1\otimes 2^+_{ms})_1\parallel{\cal M}(E1)\parallel 0^+_1\rangle\gg \langle (3^-_1\otimes 2^+_1)_1\parallel{\cal M}(E1)\parallel 0^+_1\rangle$.

The last statement seems to be in disagreement with the analytical result presented in
\cite{Smirnova} (Table 3). A possible reason is connected with a different proton--neutron structure
of the shell model E1 transition operator and the two--body term in the $sdf$ IBM--2 E1 transition operator.
Roughly speaking, the proton and neutron parts of the shell model E1 transition operator
has opposite signs. Only a small part of this operator proportinal to $(N-Z)/A$, which appears because of substraction
of the center of mass motion, contains proton and neutron contributions with the same signs. However,
the two--body term in the $sdf$ IBM--2 E1 transition operator used in \cite{Smirnova} contains the proton
and neutron parts with the same sign. As a consequence, the value of B(E1;$1^-_{ms}\rightarrow 0^+_1$),  to which
only a two--body term contributes, becomes small and if $\eta\equiv e_{\nu}/e_{\pi}$=1, i.e.,
a two--body term becomes the $F$--spin vector, B(E1;$1^-_{ms}\rightarrow 0^+_1$)=0.

In this paper, we did not calculate the matrix element $\langle 3^-_1 \parallel {\cal M}(E1)\parallel 2^+_{ms}\rangle$.
However, if we take expression (\ref{mixed-eq}) as an approximation for the operator, which produces the
$2^+_{ms}$ state acting on the ground state, then the proton and the neutron contributions to this
matrix element will have the same sign and for this reason
$|\langle 3^-_1\parallel {\cal M}(E1)\parallel 2^+_{ms}\rangle |$ will be larger than
$|\langle 3^-_1\parallel {\cal M}(E1)\parallel 2^+_1\rangle |$.

\section{RPA based microscopic consideration}

The aim of this section is to calculate the dipole transitional matrix element
$\langle 1^-_1\parallel{\cal M}(E1)\parallel 0^+_1\rangle$,
basing on expressions (\ref{1-eq}), (\ref{3-eq}) and (\ref{5-eq}) for the state vectors
$|2^+_1\rangle$, $|3^-_1\rangle$ and $|1^-_1\rangle$ and on the RPA
approximation for the ground state. Thus, in the calculations below it is assumed that the wave vectors
of the $2^+_1$, $3^-_1$ and $1^-_1$ states are
\begin{equation}
\label{22-eq}
|2^+_1,\mu\rangle =\tilde{{\cal N}}_{2^+_1}\hat{Q}_{2\mu}|0^+_1 ,RPA\rangle ,
\end{equation}
\begin{equation}
\label{23-eq}
|3^-_1,\nu\rangle =\tilde{{\cal N}}_{3^-_1}\hat{Q}_{3\nu}|0^+_1 ,RPA\rangle ,
\end{equation}
and
\begin{equation}
\label{24-eq}
|1^-_1, M\rangle =\tilde{{\cal N}}_{1^-_1}\left(\hat{Q}_{2}\hat{Q}_{3}\right)_{1M}|0^+_1 ,RPA\rangle ,
\end{equation}
Since the derivation of  relations (\ref{14-eq}) and (\ref{18-eq}) is independent of a concrete structure of the $0^+_1$
state and is based only on the $Q$--phonon form of the wave vectors, the results obtained below are
consistent with relations (\ref{14-eq}) and (\ref{18-eq}).

We assume below that the ground state contains only those correlations
which are produced by the quadrupole--quadrupole and the octupole--octupole
interactions. The quadrupole and octupole interaction constants are
fixed so as to reproduce the experimental values of the
B(E2;$0^+_1\rightarrow 2^+_1)$ and the B(E3;$0^+_1\rightarrow 3^-_1)$,
respectively. The ground state correlations related to the dipole--dipole
interaction are not included. Therefore, calculating the strength of the
E1 transitions we must introduce the core polarization factor $\chi$
which takes into account a shift of a part of the E1 strength to the giant dipole
resonance
\cite{Bohr}. Mention that the RPA--type calculations of the E1 transitions
were performed in \cite{Voronov,Ponomarev1,Ponomarev2,Tsoneva}.

In our approach based on the $Q$--phonon description
the following expressions describe a microscopic structure
of the $|2^+_1\rangle$ and $|3^-_1\rangle$ state vectors
\begin{eqnarray}
\label{25-eq}
|2^+_1,\mu\rangle ={\tilde{\cal N}}_{2^+_1}\left(b^+_{2\mu}-\frac{2\sqrt{21}}{P(2)}\sum_{ss'}W_{ss'}(3)(\alpha^+_s\alpha^+_{s'})_1 b^+_3)_{2\mu}\right)|0\rangle ,
\end{eqnarray}
\begin{eqnarray}
\label{26-eq}
|3^-_1,\mu\rangle ={\tilde{\cal N}}_{3^-_1}\left(b^+_{3\mu}+\frac{2\sqrt{15}}{P(3)}\sum_{ss'}W_{ss'}(2)(\alpha^+_s\alpha^+_{s'})_1 b^+_2)_{3\mu}\right)|0\rangle ,
\end{eqnarray}
where
\begin{eqnarray}
\label{27-eq}
W_{ss'}(3)=\sum_t \langle t\parallel i^2r^2Y_2\parallel s\rangle (-1)^{j_t +j_{s'}}(u_t u_s - v_t v_s )\left\{3\quad 2\quad 1\atop j_s\quad j_{s'}\quad j_t\right\}\varphi^{(3)}_{ts'},
\end{eqnarray}
\begin{eqnarray}
\label{28-eq}
W_{ss'}(2)=\sum_t \langle t\parallel i^3r^3Y_3\parallel s\rangle (-1)^{j_t +j_{s'}}(u_t u_s - v_t v_s )\left\{2\quad 3\quad 1\atop j_s\quad j_{s'}\quad j_t\right\}\varphi^{(2)}_{ts'},
\end{eqnarray}
\begin{eqnarray}
\label{29-eq}
P(\lambda )=\sum_{ss'}\langle s\parallel i^{\lambda}r^{\lambda}Y_{\lambda}\parallel s'\rangle (u_s v_{s'}+u_{s'} v_s )\left(\psi^{(\lambda )}_{ss'} +\varphi^{(\lambda )}_{ss'}\right).
\end{eqnarray}
In the above expressions summation is performed over both proton and
neutron single particle states. In (\ref{25-eq}) and (\ref{26-eq})
$b^+_{2\mu}$ and $b^+_{3\mu}$ are the creation operators of
the most collective quadrupole and
octupole RPA phonons corresponding to the first roots of the RPA secular
equation; $\alpha^+_s$ is the quasiparticle creation operator;
$\psi^{(\lambda )}_{ss'}$ and $\varphi^{(\lambda )}_{ss'}$ are the RPA amplitudes
describing a microscopic structure of the most collective phonons
$b^+_{2\mu}$ and $b^+_{2\mu}$; $u, and v$ are the coefficients of the
Bogoliubov transformation; ${\tilde{\cal N}}_{\lambda_1}$ is the
normalization factor. In (\ref{25-eq}) and (\ref{26-eq}) dipole
excitations are presented by the two--quasiparticle  and not by the
phonon operators because dipole--dipole correlations in the ground state
are not taken into account. Of course,  summation is performed over all
$I^{\pi}=1^-$ two--quasiparticle states. Contributions of the second
terms in
(\ref{25-eq}) and (\ref{26-eq}) to the total norm are of the order of
1$\%$ or less.

The wave vector of the $1^-_1$ state contains the following
components: $(b^+_2b^+_3)_1|0\rangle$,$\quad$ $\sum_{ss'}W_{ss'}(3)(\alpha^+_s\alpha^+_{s'})_1|0\rangle$
and $\sum_{ss'}W_{ss'}(2)(\alpha^+_s\alpha^+_{s'})_1|0\rangle$. Among them
the first one gives the main contribution to the norm, which is close
to 100$\%$. There are also other components in
the wave vectors of the $2^+_1$, $3^-_1$ and $1^-_1$ states.  However, they
are small and unimportant for  calculations of the E1 transitions. A similar conclusion about a microscopic structure
of the $1^-_1$ state was obtained in \cite{Ponomarev1}.

The components of the wave vectors of the $2^+_1$, $3^-_1$ and $1^-_1$ states, containing
operators creating
two--quasiparticle $1^-$ states, appear because the quadrupole and the octupole multipole
operators representing the corresponding state vectors according to (\ref{1-eq}), (\ref{3-eq}) and (\ref{5-eq}),
are not exhausted by the one--boson term expressed in collective bosons only.
In the calculations based on the Quasiparticle--Phonon model \cite{Soloviev,Grinberg,Voronov,Ponomarev1,Ponomarev2,Tsoneva}
this admixture  is generated
by the quasiparticle--phonon coupling term which is produced by the terms in
$Q_2$ and $Q_3$ additional to the one -- boson term. However, the weight of this contribution in the norm
of the eigenstate is similar in both the approaches.
For the reduced matrix element $\langle 1^-_1\parallel {\cal M}(E1)\parallel 0^+_1\rangle$
we obtain the following expression:
\begin{eqnarray}
\label{30-eq}
\langle 1^-_1\parallel {\cal M}(E1)\parallel 0^+_1\rangle =\frac{e}{2}(1+\chi ) \sqrt{ \frac{4\pi}{3}}\left((1+\frac{N-Z}{A})B^{(\pi)} - (1-\frac{N-Z}{A})B^{(\nu)}\right),
\end{eqnarray}
where
\begin{eqnarray}
\label{31-eq}
B^{(\pi)}=-\sqrt{\frac{35}{(1+P^{(\nu)2}(2)/P^{(\pi)2}(2))(1+P^{(\nu)2}(3)/P^{(\pi)2}(3))}}\left(Z^{(\pi )} + 2\frac{T^{(\pi)}(3)}{P^{(\pi)}(3)} + 2\frac{T^{(\pi)}(2)}{P^{(\pi)}(2)} \right),
\end{eqnarray}
\begin{eqnarray}
\label{32-eq}
B^{(\nu)}=-\sqrt{ \frac{35}{(1+P^{(\pi)2}(2)/P^{(\nu)2}(2))(1+P^{(\pi)2}(3)/P^{(\nu)2}(3))}}\left(Z^{(\nu )} + 2\frac{T^{(\nu)}(3)}{P^{(\nu)}(3)} + 2\frac{T^{(\nu)}(2)}{P^{(\nu)}(2)} \right),
\end{eqnarray}
\begin{eqnarray}
\label{33-eq}
Z^{(\tau)}=\sum_{s,s',t\in\tau }\langle s\parallel irY_1\parallel s'\rangle (-1)^{j_s + j_t}(u_s u_{s'}-v_s v_{s'})\left\{2\quad 3\quad 1\atop j_s\quad j_{s'}\quad j_t\right\}\left(\psi^{(3)}_{st}\psi^{(2)}_{s't} + \varphi^{(3)}_{st}\varphi^{(2)}_{s't}\right),
\end{eqnarray}
\begin{eqnarray}
\label{34-eq}
T^{(\tau)}(3)=-\sum_{ss'\in\tau}\langle s\parallel irY_1\parallel s'\rangle (u_s v_{s'}+u_{s'}v_s )W^{(\tau)}_{ss'}(2),
\end{eqnarray}
\begin{eqnarray}
\label{35-eq}
T^{(\tau)}(2)=-\sum_{ss'\in\tau}\langle s\parallel irY_1\parallel s'\rangle (u_s v_{s'}+u_{s'}v_s )W^{(\tau)}_{ss'}(3).
\end{eqnarray}
In (\ref{33-eq}), (\ref{34-eq}) and (\ref{35-eq}) $\tau =\pi$ or $\nu$.
The quantities $P^{(\tau)}(\lambda )$ and the matrices $W^{(\tau)}_{ss'}(\lambda )$
are determined by the expressions analogous to (\ref{29-eq}) and
(\ref{27-eq} -- \ref{28-eq}), respectively, though with summation over
proton or neutron single particle states only.

The results of the calculations of the electric dipole transitional
matrix element are presented in Figures 1 and 2 and in  Tables 1--9.
Here besides the total
calculated dipole transitional matrix element
$|\langle 1^-_1\parallel{\cal M}(E1)\parallel0^+_1\rangle |_{total}$
including all
contributions  the results
obtained without inclusion of the contribution coming
from the $I^{\pi}=1^-$ two--quasiparticle admixture denoted by
$|\langle 1^-_1\parallel{\cal M}(E1)\parallel0^+_1\rangle |_{T=0}$ are also shown.

The results presented in Tables 1 and 2 and in Figs. 1 and 2 show that in Cd, Sn, Ba, Ce and partly in Nd and Sm isotopes
the experimental data are between  the results of calculations obtained with and without a contribution
of an admixture of a dipole two--quasiparticle component to the $1^-_1$ state. However, in many cases
the matrix element $|\langle 0^+_1\parallel {\cal M}\parallel 1^-_1\rangle |_{T=0}$, which includes only a collective contribution,
is closer to the experimental data than the total matrix element, which includes both collective and
two -- quasiparticle contributions. Nevertheless, we can see that the two -- quasiparticle admixture to the
collective quadrupole -- octupole two -- phonon component of the $1^-_1$ state should be taken into account
to improve agreement with the experimental data. In the present calculations, this contribution of the
two -- quasiparticle component is overestimated. However, since the weight of this component is smaller than 1$\%$ it is difficult
to expect such an accuracy from the microscopic calculations.
We can conclude that the two--phonon
component $(|2^+_1\rangle\otimes |3^-_1\rangle )_{1^-}$,
which gives the main contribution to the norm of the $|1^-_1\rangle$ state,
determines an excitation energy and the
E2 and E3 decay properties of the $|1^-_1\rangle$ state. However, for description of the
electric dipole transitions $0^+_1\rightarrow 1^-_1$ it is necessary
to take into account also an admixture of the dipole two--quasiparticle component.

As it is seen from expressions (\ref{25-eq}) and (\ref{26-eq}), the  components containing two--quasiparticles coupled
 to angular momentum $I$=1 contribute also to the structure of the $2^+_1$ and $3^-_1$ states. Since the microscopic E1
 transitional operator contains in addition to the quadrupole -- octupole phonon term
 a term changing the number of quasiparticles by two units, these components contribute to the
 $\langle 3^-_1\parallel{\cal M}(E1)\parallel 2^+_1\rangle$ transition matrix element.

Both sets of the results obtained with and without inclusion of the contribution of the two--quasiparticle
component of the $1^-_1$ state
reproduce also an experimental
$A$--dependence of the reduced matrix element.
Only in $^{146}$Nd and $^{148}$Sm a depth of the minimum in the $A$--dependence of
$\langle 0^+_1\parallel {\cal M}\parallel 1^-_1\rangle$ is significantly larger than in the
calculated results.
As it was noted in Introduction
this matrix element has a minimum when the number of neutrons is equal to the magic number
plus(minus) four neutrons. In Tables 3--8 the calculated ratios of the matrix elements
$\langle 0^+_1\parallel {\cal M}\parallel 1^-_1\rangle |_A$/$\langle 0^+_1\parallel {\cal M}\parallel 1^-_1\rangle |_{semi magic}$
are compared with the experimental ones. It is seen that in the case of Cd, Sn, Ba and
Ce isotopes our calculations reproduce an $A$--dependence of the reduced dipole matrix element.
However, in Nd and Sm isotopes experimental data have a deeper minimum than the calculated
values. Calculations rather show a delay in an increase of the absolute value of the matrix
element with increasing a number of the valence particles or holes.

In the case of the Nd and Sm isotopes the results of the microscopic calculations deviate from
the picture which follows from the analysis of an $A$--dependence of
$|\langle 0^+_1\parallel {\cal M}\parallel 1^-_1\rangle |$ in the Interacting Boson Model,
where it is clearly seen that the reduced matrix element $|\langle 0^+_1\parallel {\cal M}\parallel 1^-_1\rangle |$
decreases if the number of valence protons is fixed, but the number of  valence neutrons
increases from zero. In the microscopic calculations, a proton contribution to the E1 transition
matrix element changes when the number of valence neutrons varies although the number of
valence protons continue to be fixed. This fact is illustrated by the results presented in
Table 9 where the quantities $B^{({\pi})}$ and $B^{({\nu})}$ determined by
Eqs. (\ref{31-eq}--\ref{32-eq}) are shown for the Nd isotopes.

The nature of the electric dipole two--quasiparticle component admixed to the wave function
of the $1^-_1$ state is an interesting question. Since this component is a mixture of
many spherical two--quasiparticle states, we can consider it formally as the $p$--boson introduced in
\cite{Iachello}. If we separate this component from the wave function of the $1^-_1$ state,
change its normalization coefficient so as to obtain a wave function with the norm equal
to one and calculate the E1 reduced matrix element between this artificially constructed state
and the ground state, we obtain the value of the order of 0.6--0.7 $e\cdot$fm. For
comparison, for a pure alpha--cluster state we obtain approximately 6 $e\cdot$fm.
For the 1p--1h (1h$_{11/2}$ 1g$^{-1}_{9/2}$)$_{1^-}$ proton state this matrix element is equal
to 1.7 $e\cdot$fm. So, it is difficult to do a definite conclusion
about the nature of this two--quasiparticle admixture.

Concluding this section we should like to stress that any microscopic consideration of the enhanced E1 transitions
cannot be performed in a harmonic approximation, i.e., cannot be carried out in the framework of the pure RPA and requires
from the beginning an inclusion of the anharmonic effects. Indeed, in the pure RPA calculations
only those parts of the one--body fermion operators are taken into account which change the number of quasiparticles
by two units. The terms of the form $\alpha^+_s\alpha_t$ are neglected. The only exception is  the
part of the nuclear Hamiltonian which describes noninteracting quasiparticles because of its leading role.
However, just the term of the form $\alpha^+_s\alpha_t$ describes $3^-_1\rightarrow 2^+_1$ E1 transition.
Therefore, the terms of this kind in the $Q_2$ and the $Q_3$ operators should be included into
consideration. However, this inclusion leads  to appearance of the anharmonic terms in the Hamiltonian
since the latter contains quadrupole -- quadrupole and octupole -- octupole interactions. An inclusion
into the operators $Q_2$, $Q_3$ and ${\cal M}(E1)$ of the terms proportional to $\alpha^+\alpha$ is necessary to conserve
commutativity of these operators. This conclusive remark shows that a comparison of the present microscopic
consideration, which from the beginning contains anharmonic effects, with the U(5) limit of IBM
is difficult.

\section{Summary}

In conclusion,  based on the $Q$--phonon representation of the wave vectors of
the low--lying collective states we have derived the relations between different reduced matrix elements
of the E1 transition operator. These relations explain qualitatively the experimentally observed correlations
among data on E1 transitions.

Using the RPA approximation for the ground state vector and the $Q$ -- phonon form of the $1^-_1$ state
we performed microscopic calculations of the reduced matrix element of the E1 transition operator between
$|0^+_1\rangle$ and $|1^-_1\rangle$ states. It is shown that the two--quasiparticle
component of the wave vector of the $|1^-_1\rangle$ state should be taken into account to achieve an agreement with
the experimental data
in spite of a small contribution of this component to the norm
of the $|1^-_1\rangle$ state.

\section{Acknowledgment}

The authors are grateful to Profs.  P. von Brentano, J. Jolie, A. Richter, W. Scheid and A. Zilges and
Dr.V. Ponomarev for
very useful discussions.
This work was supported in part by
RFBR (Moscow), grant 04--02--17376.

\begin{figure}
\caption{\label{fig:CdSnBa}
Absolute values of the reduced matrix elements for the E1 $0^+_1\rightarrow 1^-_1$
transitions in Cd, Sn and Ba isotopes calculated with an inclusion of all contributions
(dashed line), without inclusion of the contribution coming from the two--quasiparticle admixture
(dot line) and the experimental data (solid line).}
\end{figure}

\begin{figure}
\caption{\label{fig:CeNdSm}
The same as in Fig. 1, but for Ce, Nd and Sm isotopes.}
\end{figure}

\begin{table}
\caption{\label{tab:table1}
The experimental (exp) and calculated electric dipole transition matrix elements for Cd, Sn and Ba
isotopes obtained including all contributions (total) and without contribution of the
two--quasiparticle admixture to the $1^-_1$ state (T=0) (in units $e\cdot fm$).
}
\begin{ruledtabular}
\begin{tabular}{cccc}
Nucleus  &\hspace*{0.5cm} $|\langle 1^-_1\parallel{\cal M}(E1)\parallel0^+_1\rangle |_{total}$\hspace*{0.5cm} & $|\langle 1^-_1\parallel{\cal M}(E1)\parallel0^+_1\rangle |_{T=0}$\hspace*{0.5cm} & $|\langle 1^-_1\parallel{\cal M}(E1)\parallel0^+_1\rangle |_{exp}$ \\
\hline
$^{108}$Cd & 0.140 & 0.018 & 0.050 \\
$^{110}$Cd & 0.119 & 0.024 & 0.048 \\
$^{112}$Cd & 0.115 & 0.026 & 0.041 \\
$^{114}$Cd & 0.114 & 0.031 & 0.044 \\
$^{116}$Cd & 0.109 & 0.036 & 0.034 \\
$^{116}$Sn & 0.170 & 0.066 & 0.081 \\
$^{118}$Sn & 0.173 & 0.073 & 0.085 \\
$^{120}$Sn & 0.169 & 0.076 & 0.087  \\
$^{122}$Sn & 0.156 & 0.073 & 0.085 \\
$^{124}$Sn & 0.134 & 0.063 & 0.078 \\
$^{134}$Ba & 0.086 & 0.019 & 0.048 \\
$^{136}$Ba & 0.113 & 0.040 & 0.071 \\
$^{138}$Ba & 0.160 & 0.082 & 0.114 \\
$^{140}$Ba & 0.152 & 0.074 & --    \\
$^{142}$Ba & 0.153 & 0.078 & --     \\
$^{144}$Ba & 0.158 & 0.079 & --    \\
\end{tabular}
\end{ruledtabular}
\end{table}

\newpage
\begin{table}
\caption{\label{tab:table2}
The same as in Table 1   but for Ce, Nd and Sm isotopes.
}
\begin{ruledtabular}
\begin{tabular}{cccc}
Nucleus  & \hspace*{0.5cm}$|\langle 1^-_1\parallel{\cal M}(E1)\parallel0^+_1\rangle |_{total}$ & \hspace*{0.5cm}$|\langle 1^-_1\parallel{\cal M}(E1)\parallel0^+_1\rangle |_{T=0}$ & \hspace*{0.5cm}$|\langle 1^-_1\parallel{\cal M}(E1)\parallel0^+_1\rangle |_{exp}$ \\
\hline
$^{140}$Ce & 0.179 & 0.099 & 0.129  \\
$^{142}$Ce & 0.170 & 0.092 & 0.108  \\
$^{144}$Ce & 0.190 & 0.114 & --   \\
$^{146}$Ce & 0.209 & 0.134 & --  \\
$^{142}$Nd & 0.189 & 0.108 & 0.128 \\
$^{144}$Nd & 0.181 & 0.101 & 0.098  \\
$^{146}$Nd & 0.187 & 0.111 & 0.071  \\
$^{148}$Nd & 0.223 & 0.141 & 0.119  \\
$^{144}$Sm & 0.193 & 0.115 & 0.140  \\
$^{146}$Sm & 0.177 & 0.103 & --  \\
$^{148}$Sm & 0.190 & 0.116 & 0.052  \\
$^{150}$Sm & 0.212 & 0.132 & 0.099  \\
\end{tabular}
\end{ruledtabular}
\end{table}

\newpage
\begin{table}
\caption{\label{tab:table5}
The experimental and calculated absolute values of the reduced matrix element
$\langle 0^+_1\parallel{\cal M}(E1)\parallel 1^-_1\rangle$ in Cd isotopes given in units of $|\langle 0^+_1\parallel{\cal M}(E1)\parallel 1^-_1\rangle |$.}
\begin{ruledtabular}
\begin{tabular}{cccccc}
Nucleus & \hspace*{0.5cm}$^{108}$Cd & \hspace*{0.5cm}$^{110}$Cd & \hspace*{0.5cm}$^{112}$Cd & \hspace*{0.5cm}$^{114}$Cd & \hspace*{0.5cm}$^{116}$Cd \\
\hline
Exp.& 1 & 1.0 & 0.80 & 0.90 & 0.70 \\
Calc. & 1 & 0.85 & 0.82 & 0.82 & 0.78 \\
\end{tabular}
\end{ruledtabular}
\end{table}

\begin{table}
\caption{\label{tab:table6}
The experimental and calculated absolute values of the reduced matrix element
$\langle 0^+_1\parallel{\cal M}(E1)\parallel 1^-_1\rangle$ in Sn isotopes given in units of
$|\langle 0^+_1\parallel{\cal M}(E1)\parallel 1^-_1\rangle |$ for $^{116}$Sn.}
\begin{ruledtabular}
\begin{tabular}{cccccc}
Nucleus & \hspace*{0.5cm}$^{116}$Sn & \hspace*{0.5cm}$^{118}$Sn & \hspace*{0.5cm}$^{120}$Sn & \hspace*{0.5cm}$^{122}$Sn & \hspace*{0.5cm}$^{124}$Sn \\
\hline
Exp.& 1 & 1.0 & 1.06 & 1.0 & 0.94 \\
Calc. & 1 & 1.02 & 0.99 & 0.91 & 0.78 \\
\end{tabular}
\end{ruledtabular}
\end{table}

\begin{table}
\caption{\label{tab:table7}
The experimental and calculated absolute values of the reduced matrix element
$\langle 0^+_1\parallel{\cal M}(E1)\parallel 1^-_1\rangle$ in Ba isotopes given in units of
$|\langle 0^+_1\parallel{\cal M}(E1)\parallel 1^-_1\rangle |$ for $^{138}$Ba.}
\begin{ruledtabular}
\begin{tabular}{cccccc}
Nucleus & \hspace*{0.5cm}$^{134}$Ba & \hspace*{0.5cm}$^{136}$Ba & \hspace*{0.5cm}$^{138}$Ba & \hspace*{0.5cm}$^{140}$Ba & \hspace*{0.5cm}$^{142}$Ba \\
\hline
Exp.& 0.43 & 0.61 & 1 & --- & --- \\
Calc. & 0.53 & 0.70 & 1 & 0.94 & 0.95 \\
\end{tabular}
\end{ruledtabular}
\end{table}

\begin{table}
\caption{\label{tab:table8}
The experimental and calculated absolute values of the reduced matrix element
$\langle 0^+_1\parallel{\cal M}(E1)\parallel 1^-_1\rangle$ in Ce isotopes given in units of
$|\langle 0^+_1\parallel{\cal M}(E1)\parallel 1^-_1\rangle |$ for $^{140}$Ce.}
\begin{ruledtabular}
\begin{tabular}{ccccc}
Nucleus & \hspace*{0.5cm}$^{140}$Ce & \hspace*{0.5cm}$^{142}$Ce & \hspace*{0.5cm}$^{144}$Ce & \hspace*{0.5cm}$^{146}$Ce  \\
\hline
Exp.& 1 & 0.85 & --- & --- \\
Calc. & 1 & 0.94 & 1.06 & 1.16  \\
\end{tabular}
\end{ruledtabular}
\end{table}

\begin{table}
\caption{\label{tab:table9}
The experimental and calculated absolute values of the reduced matrix element
$\langle 0^+_1\parallel{\cal M}(E1)\parallel 1^-_1\rangle$ in Nd isotopes given in units of
$|\langle 0^+_1\parallel{\cal M}(E1)\parallel 1^-_1\rangle |$ for $^{142}$Nd.}
\begin{ruledtabular}
\begin{tabular}{ccccc}
Nucleus & \hspace*{0.5cm}$^{142}$Nd & \hspace*{0.5cm}$^{144}$Nd & \hspace*{0.5cm}$^{146}$Nd & \hspace*{0.5cm}$^{148}$Nd  \\
\hline
Exp.& 1 & 0.77 & 0.56 & 0.92 \\
Calc. & 1 & 0.96 & 0.99 & 1.18  \\
\end{tabular}
\end{ruledtabular}
\end{table}

\begin{table}
\caption{\label{tab:table10}
The experimental and calculated absolute values of the reduced matrix element
$\langle 0^+_1\parallel{\cal M}(E1)\parallel 1^-_1\rangle$ in Sm isotopes given in units of
$|\langle 0^+_1\parallel{\cal M}(E1)\parallel 1^-_1\rangle |$ for $^{144}$Sm.}
\begin{ruledtabular}
\begin{tabular}{ccccc}
Nucleus & \hspace*{0.5cm}$^{144}$Sm & \hspace*{0.5cm}$^{146}$Sm & \hspace*{0.5cm}$^{148}$Sm & \hspace*{0.5cm}$^{150}$Sm  \\
\hline
Exp.& 1 & --- & 0.38 & 0.69 \\
Calc. & 1 & 0.91 & 0.98 & 1.10  \\
\end{tabular}
\end{ruledtabular}
\end{table}

\begin{table}
\caption{\label{tab:table11}
Dependence of the calculated proton $B^{(\pi)}$ and  neutron $B^{(\nu)}$ contributions
to the reduced matrix element $\langle 0^+_1\parallel{\cal M}(E1)\parallel 1^-_1\rangle$
on the number of neutrons in Nd isotopes. For definition of the quantities $B^{(\pi)}$ and
$B^{(\nu)}$ see (\protect\ref{30-eq}--\protect\ref{35-eq}).}
\begin{ruledtabular}
\begin{tabular}{ccc}
Nucleus & \hspace*{0.5cm}$B^{(\pi)}$ (in fm) & \hspace*{0.5cm}$B^{(\nu)}$ (in fm) \\
\hline
$^{142}$Nd$_{82}$ & 0.78 & 0.17 \\
$^{144}$Nd$_{84}$ & 0.80 & 0.21 \\
$^{146}$Nd$_{86}$ & 0.87 & 0.26 \\
$^{148}$Nd$_{88}$ & 1.05 & 0.32 \\
\end{tabular}
\end{ruledtabular}
\end{table}

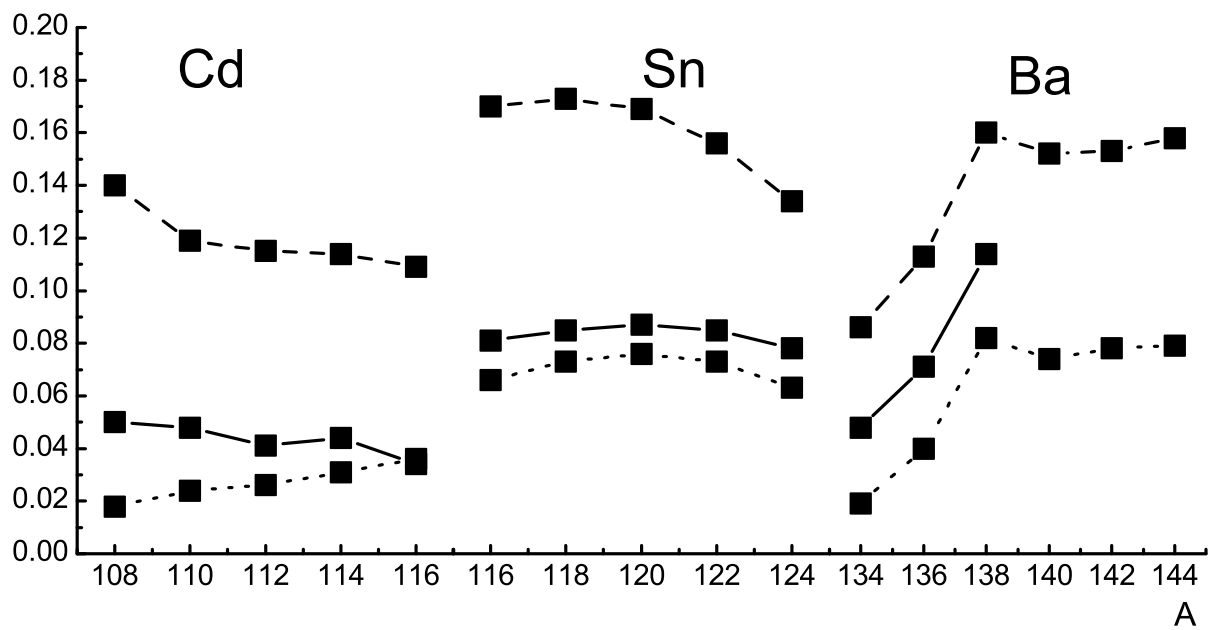
\end{document}